\documentclass[useAMS,usenatbib]{mn2e}
\usepackage{rotating}
\usepackage{journals}
\usepackage{graphics,graphicx}

\newcommand\msun {M$_{\odot}$}
\newcommand\gtrsim{\mathrel{\hbox{\rlap{\hbox{\lower4pt\hbox{$\sim$}}}\hbox{$>$}}}}
\def\approxgt{\ifmmode \rlap{$>$}{}_{{}_{{}_{\textstyle\sim}}} \else%
$\rlap{$>$}{}_{{}_{{}_{\textstyle\sim}}}$\fi} 
\def\approxlt{\ifmmode \rlap{$<$}{}_{{}_{{}_{\textstyle\sim}}} \else%
$\rlap{$<$}{}_{{}_{{}_{\textstyle\sim}}}$\fi}

\def\farcs{\hbox{$.\!\!^{\prime\prime}$}}

\def\arcmin{\hbox{$^\prime$}}
\def\arcsec{\hbox{$^{\prime\prime}$}}

\def\flx{erg cm$^{-2}$ s$^{-1}$}
\def\lum{erg s$^{-1}$}
\def\xmm{XMM-{\it Newton}}
\def\chan{{\it Chandra}}
\def\src{SWIFT~J1749.4--2807}
\def\swift{{\it Swift}}

\LARGE \normalsize \title[Candidate NIR counterpart to \src ]
{\chan\, X-ray and {\it Gemini} near--infrared observations of the eclipsing msec pulsar \src\, in quiescence}
\author[Jonker et al.]  {Peter G.~Jonker$^{1,2,3}$,
Manuel A.P.~Torres$^{1,2}$,  Danny Steeghs$^{4}$, Deepto~Chakrabarty$^5$\\ 
$^1$SRON, Netherlands Institute for Space Research,
Sorbonnelaan 2, 3584~CA, Utrecht, The Netherlands\\ 
$^2$Harvard--Smithsonian Center for Astrophysics, 60 Garden Street, 
Cambridge, MA~02138, U.S.A.\\
$^3$Department of Astrophysics/IMAPP, Radboud University Nijmegen, PO Box 
9010, 6500 GL Nijmegen, the Netherlands\\
$^4$Astronomy and Astrophysics, Department of Physics, University
of Warwick, Coventry, CV4~7AL\\
$^5$Department of Physics and Kavli Institute for Astrophysics and
Space Research, Massachusetts Institute
of Technology, Cambridge, MA 02139\\ U.S.A.\\
}
\begin{document}

\maketitle

\begin{abstract} \noindent We report on \chan\, X-ray and {\it
    Gemini--North} near--infrared $K$--band observations of the
  eclipsing accretion--powered millisecond X--ray pulsar \src\, in
  quiescence.  Using the \chan\, observation we derive a source
  position of Right Ascencion: 17:49:31.73 and
  Declination:-28:08:05.08. The position is accurate to 0\farcs6 (90
  per cent confidence). We find one source at a magnitude
  $K$=18.44$\pm$0.03 with a position fully consistent with the
  accurate \chan~ X--ray localisation and a second source at
  $K$=19.2$\pm$0.1 that falls close to the edge of the error circle in
  the deep $K$--band images. The presence of a few weaker sources as
  suggested by previous $H$--band observations presented in the
  literature cannot the ruled out. There is marginal evidence that the
  brighter of the these two sources is variable. Follow-up
  spectroscopy of this potential counterpart will show if this source
  is the true counterpart to \src. If so, baring the presence of
  complicating effects such as heating of the mass-donor star, it
  would allow for the mass of the neutron star to be measured through the
  measurement of periodic radial velocity variations.

 \end{abstract}


\section{Introduction} 

Neutron stars provide a laboratory to test the behaviour of matter under
physical conditions that are unattainable on Earth.  The description of the
relations between pressure and density of matter under the extreme
conditions encountered in neutron stars (the equation of state; EoS) is one
of the ultimate goals of the study of neutron stars already for 40 years
now.

An excellent way to constrain proposed EoS is by finding massive
neutron stars, as each of the EoS predicts a maximum mass for a
neutron star above which it will collapse into a black hole
(\citealt{2001ApJ...550..426L}).  Neutron stars more massive than the
canonical value of 1.4 M$_\odot$ can be found in X--ray binaries, as
in those systems the neutron stars can accrete a substantial amount of
mass from its binary companion. There is strong evidence that
millisecond pulsars evolve out of low--mass X--ray binaries (LMXBs),
see for instance \citet{1998Natur.394..344W},
\citet{1998Natur.394..346C} and \citet{2009Sci...324.1411A}. Whereas
the average mass of neutron stars that accreted substantially is
$1.5\pm0.2$ M$_\odot$ (\citealt{2012arXiv1201.1006O}, see also
\citealt{2011A&A...527A..83Z}), there is a sample of millisecond
pulsars where high neutron stars masses have been measured. For
instance, the mass of the neutron star in FIRST~J102347.6$+$003841 has
recently been determined to be 1.71$\pm0.16$ \msun\,
(\citealt{2012arXiv1207.5670D}) and the millisecond pulsar
PSR~J0751+1807 has a mass of 2.1$\pm$0.2
\msun\,(\citealt{2005ApJ...634.1242N}). The most accurately measured
mass of a massive neutron star is that in PSR~J1614$-$2230. It has a
mass of 1.97$\pm$0.04 \msun\, (\citealt{2010Natur.467.1081D}).
However, the maximum neutron star mass may be higher still. For
instance, the best estimate of the Black Widow pulsar mass is
2.40$\pm$0.12 \msun\, (PSR~B1957+20; \citealt{2011ApJ...728...95V}),
although the systematic uncertainty on this measurement could be
larger than the statistical one.

For dynamical mass measurements in LMXBs that host (millisecond)
pulsars, pulse timing can directly constrain the orbit of the neutron
star and thus its radial velocity semi--amplitude, $K_1$. The radial
velocity semi--amplitude of the mass donor star, $K_2$, can be
measured from optical or near--infrared observations. The ratio
between $K_1$ and $K_2$ yields the ratio between the mass of the
companion star and the neutron star, $q$. Alternatively, this $q$ can
be determined from the rotational broadening of the stellar absorption
lines ($v\sin i$).  This $v\sin i$ combined with $K_2$ gives $q$ via
$\frac{v\sin i}{K_2}=(1+q)\frac{0.49q^{2/3}}{0.6q^{2/3}+\ln
  (1+q^{1/3})}$ (e.g.~see \citealt{1986MNRAS.219..791H}). The system
inclination can be determined through modelling of the multi--colour
optical lightcurves or, in systems with favorable viewing angles such
as \src\, the X--ray eclipse duration can be used to accurately
determine the inclination (\citealt{1985MNRAS.213..129H}). As the
inclination is constrained by the geometry
(\citealt{1976ApJ...208..512C}), mass measurements in eclipsing
systems are independent of the modelling that lies behind inclinations
derived from ellipsoidal variations. In most cases the system has to
be in quiescence such that the accretion continuum is suppressed
permitting the detection of photospheric absorption features from the
mass donor. These absorption features allow one to measure $K_2$ and
$v\sin i$ from optical or near--infrared spectroscopic observations.
The eclipse duration together with $q$ gives an accurate measure of
the inclination $i$ (\citealt{1976ApJ...208..512C};
\citealt{1985MNRAS.213..129H}). $q$, $i$ and $K_2$ together give the
neutron star mass.

Recently, it was discovered that the transient \src\, found in 2006
(\citealt{2006GCN..5200....1S}; see also
\citealt{2009MNRAS.393..126W}) exhibits pulsations at 518 Hz
(\citealt{2011ApJ...727L..18A}).  Furthermore, the X--ray light curve
shows eclipses at the 8.8 hr orbital period
(\citealt{2010ApJ...717L.149M}). For the companion star to fill its
Roche lobe in the 8.8 hours orbit its mean density would imply a
spectral type of a G3~V--G5~V star if the companion is close to the
lower main-sequence.  Thus, this source holds the promise of allowing
for a model--independent neutron star mass determination.  From a type
I X-ray burst the upper limit to the distance to this source has been
determined to be 6.7$\pm$1.3 kpc (\citealt{2009MNRAS.393..126W}).

The observed interstellar hydrogen column density $N_H$ towards the
source is $\approx 3 \times10^{22}$cm$^{-2}$
(\citealt{2011A&A...525A..48F}). As it implies about 17 magnitudes of
extinction in the $V$--band (\citealt{1995A&A...293..889P}) it
precludes the search of an optical counterpart. Given that the
extinction in the $K$--band is only $\approx$11\% of that in the
$V$--band (\citealt{1989ApJ...345..245C}) it allows for the detection
of a near--infrared counterpart. \citet{2011A&A...534A..92D} searched
for a near--infrared counterpart in the 1\farcs6 \swift\, error circle
in the $H$--band. Given the crowded field, the relatively large
uncertainty on the \swift\, X--ray position and the deep observations,
\citet{2011A&A...534A..92D} found more than 40 potential counterparts.

In this Paper we report on a \chan\, observation of \src\,  in
quiescence.  This observation provides an accurate localisation. Using
this accurate position we investigate our deep {\it Gemini} $K$--band
observations obtained under excellent natural seeing conditions for
the presence of a counterpart to \src.  Throughout this paper we use
the ephemeris of \src\, given by \citet{2011ApJ...727L..18A}.

\section{Observations, analysis and results} 

\subsection{{\it Chandra} X-ray observation} 

We observed \src\, with the \chan\, satellite using the
back--illuminated S3 CCD--chip of the Advanced CCD Imaging
Spectrometer (ACIS) detector (\citealt{1997AAS...190.3404G}). The
observation started on MJD 56138.09427 (July 30, 2012, 02:15:44 UT)
and the on--source exposure time was 24.4 ksec. The observation
identification number is 13704.

To mitigate potential pile-up on the off chance that the source was
brighter than predicted we windowed the ACIS--S3 CCD such that only
half the original CCD size is read out, providing a frame time of
1.54~s of which 0.04~s is used for the CCD read out. We have
reprocessed and analysed the data using the {\it CIAO 4.3} software
employing the calibration files from the Calibration Database version
4.5. In our analysis we have selected events only if their energy
falls in the 0.3--7 keV range.  All data have been used, as background
flaring is very weak or absent.

Since, by design, the source position falls near the optical axis of
the telescope (see below), the size of the point spread function is
smaller than the ACIS pixel size. Therefore, we follow the method of
\citet{2004ApJ...610.1204L} implemented in the CIAO 4.3 tool {\sc
  acis\_process\_events} to improve the image quality of the ACIS
data.

\begin{figure}
\includegraphics[width=8cm,angle=0]{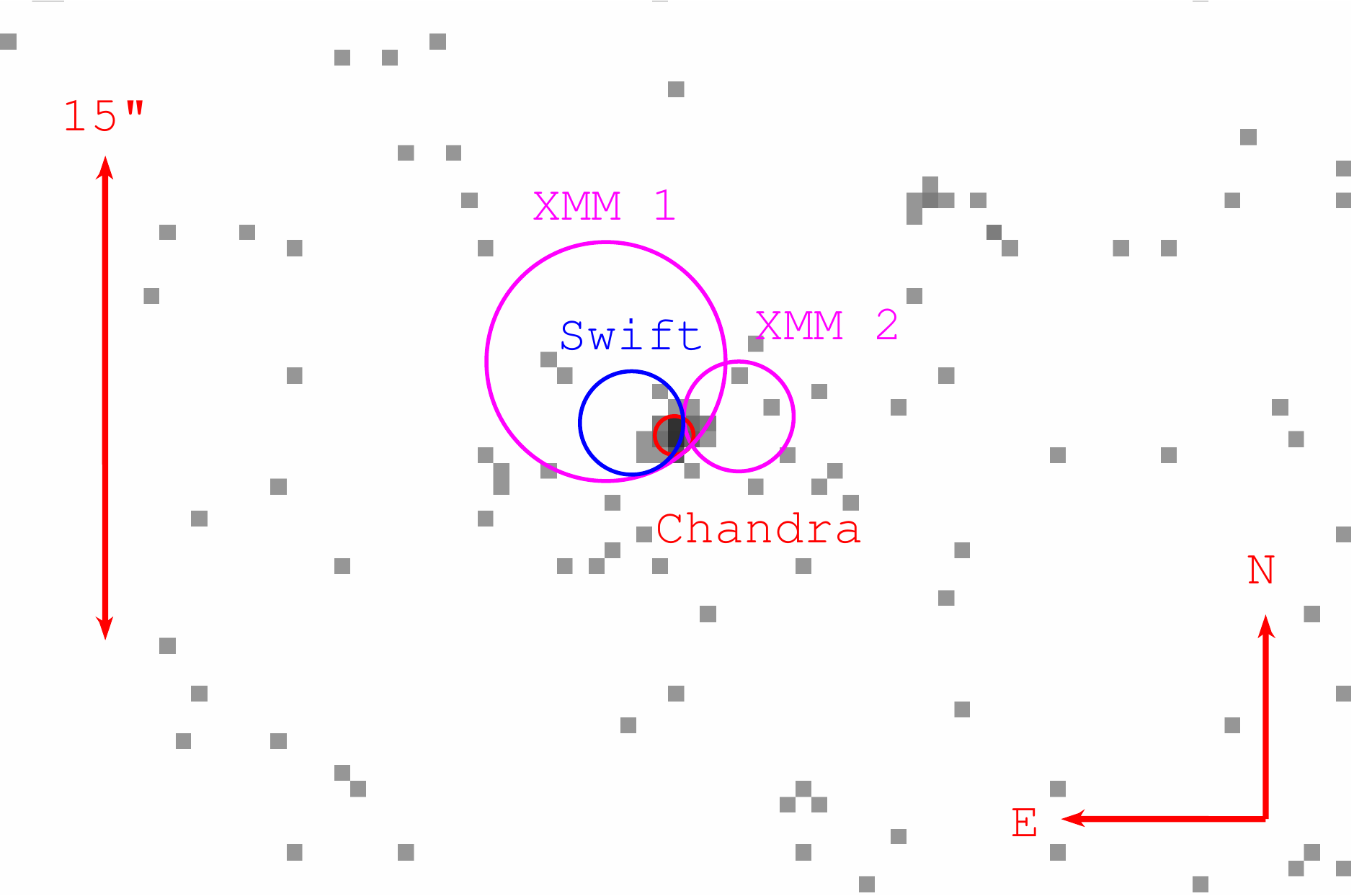}
\caption{Zoom--in around the \chan\, ACIS-S3 position of \src\,
  (indicated with the small red circle with a radius of 0\farcs6). The
  blue circle indicates the \swift\, (\citealt{2011A&A...534A..92D})
  position and the magenta circles represent the \xmm\, positions of
  \src\, given by \citet{2009MNRAS.393..126W} (XMM~1) and
  \citet{2012ApJ...756..148D} (XMM~2).}\label{ima}
\end{figure}

Using {\sc wavdetect} we detect one X--ray source within the \swift\,
(\citealt{2011A&A...534A..92D}) and \xmm\, (\citealt{2009MNRAS.393..126W})
error circles of \src\, (see Figure~\ref{ima}). The source is detected
0.2\arcmin\,off--axis on the ACIS S3 CCD. The {\sc wavdetect } J2000 source
position in decimal degrees is: right ascencion (RA): 267.38220(2) and
declination (Dec): -28.13474(1). The digit in--between brackets denotes the
{\sc wavdetect} 68 per cent confidence error on RA and Dec. The position in
sexagesimal notation is: RA: 17:49:31.73 Dec:-28:08:05.08. The errors given
above are due to the error in the source localisation on the CCD alone. The
absolute accuracy of the source position is thus determined by the
uncertainty in the boresight of the \chan\, satellite. For sources on the
ACIS-S3 CCD this uncertainty is typically 0.6\arcsec\, at 90 per cent
confidence. We tried to reduce this boresight uncertainty by investigating
if one of the other detected X-ray sources can be safely associated with a
source with a well known position but this was not the case.

We have extracted source counts from a circular region of
2\arcsec\,radius centered on the source position.  Similarly, we have
used a circular region with a radius of 50\arcsec\, on a source--free
region of the CCD to extract background counts. A point--source
aperture correction was applied to the auxiliary response file of the
source. The net, background subtracted, detected source count rate is
2.9$\times 10^{-3}$ counts s$^{-1}$.  In total after correcting for
the number of expected background photons, 61 source photons have been
detected in the energy range between 0.3--7 keV. The number of
background photons in the same area in 0.3--7 keV was less than 1.

Using {\sl xspec} version 12.6.0q (\citealt{ar1996}) we have fitted the spectra
of \src\ using Cash statistics (\citealt{1979ApJ...228..939C}) modified to
account for the subtraction of background counts, the so called
W--statistics\footnote{see
http://heasarc.gsfc.nasa.gov/docs/xanadu/xspec/manual/}.

Given that we only detected a low number of counts we decided to fix
the ${\rm N_H}$ to 3$\times 10^{22}$ cm$^{-2}$ in our spectral fits,
consistent with the values determined by \citet{2011A&A...525A..48F}
and \citet{2012ApJ...756..148D}.  We have used an absorbed powerlaw
model fit function to describe the data (the model {\sl pegpwrlw} in
{\sl xspec}). We visually inspected a plot of the count rate as a
function of heliocentric time to search for the presence of the
eclipse.  The \chan\, observed count rate is consistent with zero at
the predicted heliocentric time of the eclipse. The eclipse duration
is consistent with the duration determined by
\citet{2010ApJ...717L.149M} when the source was in outburst.  To
account for the fact that the source is eclipsed during 2~ksec
(\citealt{2010ApJ...717L.149M}), which is about 8 per cent of the
total exposure time, we decrease the effective exposure by 2~ksec when
calculating the source flux. We find a best--fit power law index of
0.6$\pm$0.4 which gives an unabsorbed X--ray flux of
$(0.8\pm0.1)\times 10^{-13}$ \flx~ in the 0.3--7 keV band and
$1.4\times 10^{-13}$ \flx~ in the 0.5--10 keV band.

\subsection{{\it Gemini} near--infrared observations }

$K$-band imaging of the field containing  \src\, was performed with the
{\it Gemini--North} telescope on the nights of June 15 and 20, 2012
under observing program  GN-2012A-Q-67. The Near Infrared Imaging and
Spectrometer (NIRI; \citealt{2003PASP..115.1388H}) instrument was used
with the f/14 camera to yield a field of view of  $51'' \times  51''$ 
with a plate scale of 0\farcs05 pixel$^{-1}$. 

The field of \src\, is crowded and care was taken to avoid as much as
possible image artifacts due to saturated stars. In particular, we
exclude from the NIRI FOV the $K\sim 8.6$ star
2MASS~J17493218--2807563.  Therefore, we both rotated the instrument
45 degrees and used a ({\it p},{\it q}=x,0)-dithering
pattern. Integration times were 10~s in other to mitigate saturation
effects from other bright stars in the field of view.  Thus the
observing sequences consisted of a 5--point dither pattern with offset
steps {\it p} of 0, 10, -5, 5 and 15\arcsec\, and we added three 10~s
exposures at each dither position. This pattern was repeated twenty
one times yielding a total of 52.5 min on source.  The image quality
varied during the observations due to the weather conditions.  We
measure a point-spread function full-width at half maximum (FWHM)
between 0\farcs22--0\farcs36 and 0\farcs23--0\farcs58 with a median
value of 0\farcs29$\pm$0\farcs03 and 0\farcs40$\pm$0\farcs07 on June
15 and 20, respectively.

The data were processed using both {\sc python} scripts and {\sc
  iraf}\footnote{\textsc {iraf} is distributed by the National Optical
  Astronomy Observatories} tasks developed and provided by the {\it Gemini}
Observatory.  Note that sky frames were not subtracted from the data. The
observed field is too crowded to allow us to build a useful sky frame
from the dithered images and there were no nearby regions on the sky
suitable to measure the local sky background.

\begin{figure} \includegraphics[width=8cm,angle=0]{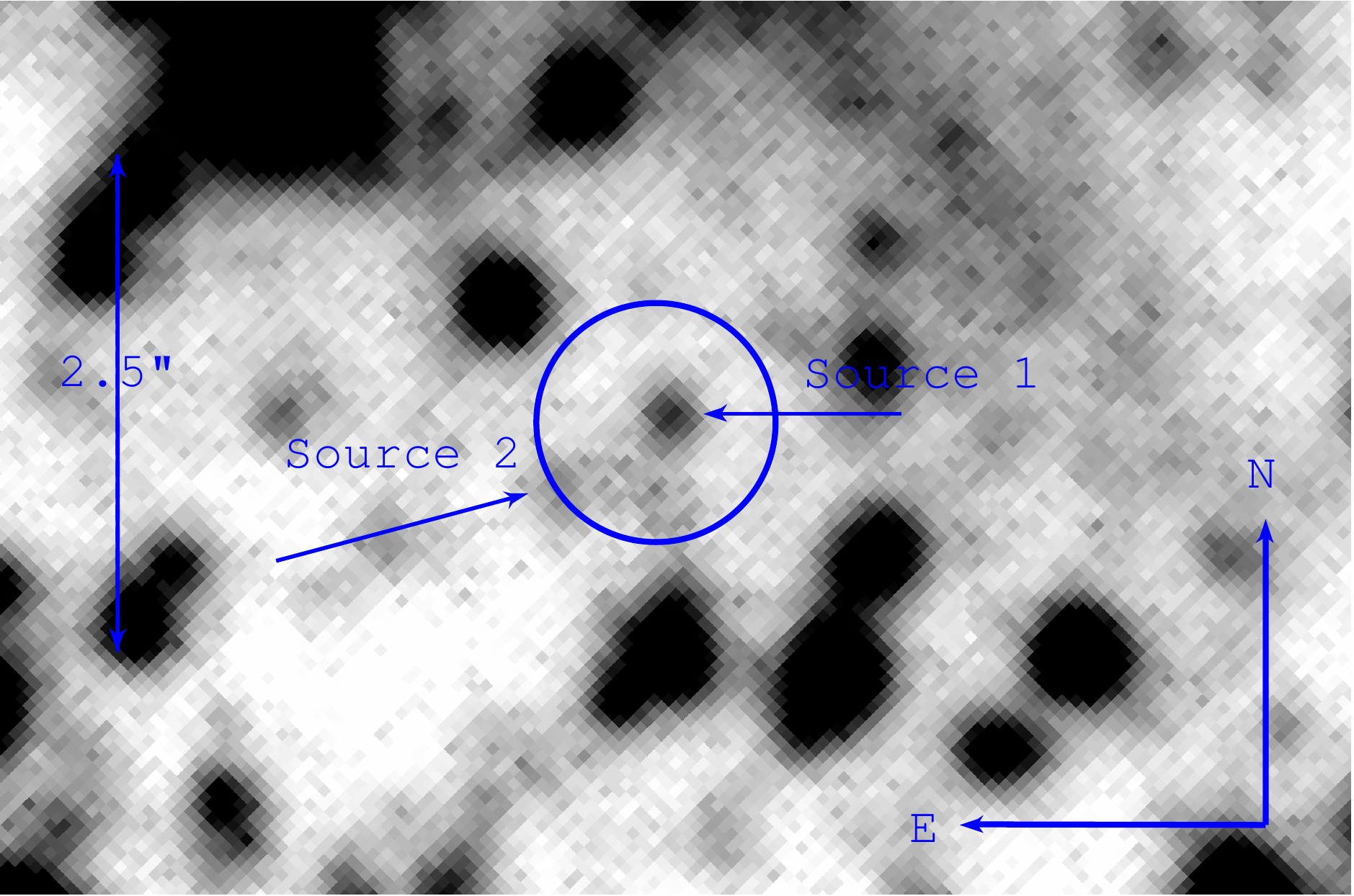}
\caption{{\it Gemini}--NIRI $K$ band finder chart obtained from combining those
images with seeing better than 0\farcs31 on June 15, 2012. The 0\farcs6 \chan~
90 per cent confidence error circle on the X--ray source position is indicated
(circle). We mark source 1 and 2, where the position of source 1 is consistent
with the X--ray position. Source 1 shows marginal evidence for variability (at
the 4~$\sigma$ level) between the observations of June 15 and June 20, 2012.
}\label{kband} \end{figure}

For the analysis, we generated images by stacking all the reduced
frames using {\it imcoadd}. In this way we obtained for the night of
June 15 seven combined images of 150~s on source each. We also
combined the images with seeing $<$0\farcs31 of the first night into
one average image.  The night of June 20 was of less good image quality
(see above) and a single combined frame was build totalling 1770~s on
source.

We improved the astrometry of the best--seeing frame using the
position of six 2MASS stars that were well separated from other stars
and which did not saturate the {\it Gemini} NIRI detector. A fit of
pixel scale in x and y, and orientation provide an root--mean--square
(rms) uncertainty of 0\farcs03 for a pixel scale of 0\farcs05 per
pixel in x and y and an orientation of 44.9$^\circ$.

In Fig.~\ref{kband} we show the finding chart of the field of \src~
using the best seeing image of June 15.  Overplotted is the 90 per
cent confidence error circle on the \chan\, X--ray position of the
source. A clear source is detected near the centre of the error circle
(star 1) and towards the south--east corner just outside the formal 90
per cent confidence error circle, lies another fainter potential
candidate counterpart (star 2).

We performed aperture photometry on the images using {\sc daophot} in
{\sc iraf} to compute the instrumental magnitudes for the objects of
interest.  Absolute flux calibration of the images was obtained
selecting isolated stars with magnitudes reported in the UKIDSS-DR6
Galactic Plane Survey (UGPS; \citealt{2008MNRAS.391..136L}). Both the
NIRI and the UKIDSS photometric systems are close representations of
the Mauna Kea Observatory (MKO) standard system. Differential
photometry was obtained on each of the averaged images to derive the
flux variability of the sources as a function of time. The photometric
results given here are with respect to the star
UGPS~J176971.32--288889.6 ($K=14.551 \pm 0.022$). The scatter in the
light curves build for this UGPS star using other bright and isolated
comparison stars shows that the differential photometry is accurate to
$\approxlt 1$ per cent.

Photometry of the two candidate counterparts on the combined best
seeing image of June 15, 2012, yields a magnitude of $K_1=18.44\pm
0.03$ and $K_2=19.20\pm 0.06$ for star 1 and star 2, respectively. The
magnitude during the observation on June 20, 2012 is
$K_1=$18.75$\pm$0.07 and $K_2=19.3\pm 0.1$ mag. So, with an average
magnitude of $K_1=18.44\pm 0.03$ on June 15, and $K_1=$18.75$\pm$0.07
on June 20, 2012, there is evidence at the 4~$\sigma$ level that this
source is variable. 

\section{Discussion}

Using \chan\, we obtained a 24.4~ksec exposure of the eclipsing
msec.~pulsar \src\, in quiescence. We detected the source at an
unabsorbed flux of $2\times 10^{-13}$ erg$^{-1}$~cm$^{-2}$~s$^{-1}$ in
the 0.5--10 keV band and we provide a position accurate to 0\farcs6 at
90 per cent confidence. At a distance upper limit of 6.7$\pm$1.3 kpc
(\citealt{2009MNRAS.393..126W}), this flux corresponds to an X--ray
luminosity of 1$\times 10^{33}$ \lum\, in the 0.5--10 keV band.  Owing
to the uncertainty in the high hydrogen column density towards the
source this number is uncertain, but the value is consistent with that
found before from \xmm\, observations in quiescence
(\citealt{2009MNRAS.393..126W}; \citealt{2012ApJ...756..148D}).  The
derived quiescent X-ray luminosity is also in agreement with the rough
correlation between quiescent X-ray luminosity and the orbital period,
although the source is on the bright end of the observed correlation
(a plot highlighting this correlation can be found in
\citealt{2011ApJ...729L..21R}). The best-fit power law index of the
X-ray spectrum is $0.6 \pm 0.4$, which is steep for quiescent neutron
stars, however, the uncertainty on this number is currently such that
it could be made consistent with more canonical values of 1.5--2
particularly if ${\rm N_H}$ is somewhat higher than the 3$\times
10^{22}$ cm$^{-2}$ that was used in the fit (see for instance the
powerlaw fit in \citealt{2012ApJ...756..148D}).

In addition to the \chan\, observation, we obtained {\it Gemini}
$K$--band images under excellent natural seeing conditions of the
field of \src. The position of one point source is consistent with the
X--ray location and a second is bordering the \chan\, error circle.
There is marginal evidence for the brighter of the two to be variable.
The observed interstellar hydrogen column density, $N_H$, towards
\src\, is $\approx 3 \times10^{22}$cm$^{-2}$
(\citealt{2011A&A...525A..48F}). This yields an extinction in the
$K$--band of 1.9 mag (\citealt{1989ApJ...345..245C}). However, as
${\rm A_V/E(B-V) = R_V}$ is known to be closer to 2.5 than 3.1 for
fields towards the bulge (\citealt{2012arXiv1208.1263N}) this implies
that $A_K$ would be 1.7 mag for ${R_V}$=2.5. 

For the companion star to fill its Roche lobe in the 8.8 hours orbit
its mean density would imply a spectral type of a G3~V--G5~V star if
the companion is close to the lower main-sequence. Note that the
spectral type of G5~V that we derive here is consistent with the broad
mass range for the companion star (0.45-0.8~\msun) that is derived by
\citet{2010ApJ...717L.149M} and \citet{2011ApJ...727L..18A} as the
mass donors in LMXBs and CVs are often under-massive for their
spectral type (\citealt{2011ApJS..194...28K}).  For a G5~V donor star,
the absolute magnitude in the $V$- and in the $K$-band is $M_V=$5.1
and $M_K=3.5$, respectively (see \citealt{2000asqu.book.....C} and
\citealt{2000asqu.book..143T}, respectively). At a distance upper
limit of 6.7 kpc the distance modulus, DM, is 14.1.  Allowing for 1.7
magnitudes of extinction in the $K$--band will mean that the companion
star has an apparent magnitude of $m_K$=DM+1.7+M$_{K}$ yielding
$m_K$=19.3 for the G5~V star. The observed $K$ band magnitude of $\sim
18.6$ is consistent with this if the source is more nearby than the
6.7 kpc upper limit (i.e.~ the source has to be at $\sim$5 kpc) and/or
if the extinction in the $K$-band is lower than 1.7 magnitudes. The
latter can be the case if part of the observed ${\rm N_H}$ is due to
ionised material for instance material local to the source.

Previous to our work, \citet{2011A&A...534A..92D} obtained $H$--band
adaptive optics observations with the Very Large Telescope. Of the 41
sources that those authors detected inside the error circle of the
\swift\, X--ray position, five fall inside the \chan\, error circle;
six if we include our source 2. Two of these six sources coincide
with the position of the two potential counterparts that we found.  In
Table ~\ref{tab-avanzo} we list the properties found by
\citet{2011A&A...534A..92D} for these six sources.

\begin{table*}
  \caption{Position and $H$--band magnitudes from \citet{2011A&A...534A..92D} 
    for the  sources close to the \chan\, 90 per cent confidence 
    error circle. The offset of the position of each of the stars with respect 
    to the centre of the \chan\, position as well as the $K$--band magnitudes 
    from this work are given. Note that the $H$--band magnitudes from
    \citet{2011A&A...534A..92D}  are
    off; approximately $-1.3$ has to be added to the reported
    magnitudes (see Discussion).}
  \label{tab-avanzo}
\begin{center}
\begin{tabular}{ccccccc}
\hline
RA & Dec & Src \# & $H$-band mag &$H$-band mag & Offset & $K$--band \\ 
   &     & D'Avanzo et al.~2011 & Aug 30, 2010 & Aug 31, 2011   & \arcsec&    mag    \\
\hline
17:49:31.730 & -28:08:04.50 & 32 & 22.75$\pm$0.11 & 22.66$\pm$0.09 & 0.58 & --\\
17:49:31.76  &  -28:08:05.39 & 41 &23.99$\pm$0.35  & 22.68$\pm$0.09 &0.53& --\\
17:49:31.720 & -28:08:05.50  & 25 &22.22$\pm$0.07 & 21.86$\pm$0.07 & 0.44& --\\
17:49:31.730 & -28:08:05.40 & 27 &22.54 $\pm$0.08 & 22.38$\pm$0.08 & 0.32 & --\\
17:49:31.72  & -28:08:05.1 & 8 & 20.81$\pm$0.05 & 20.74$\pm$0.05 & 0.11 & 18.44$\pm$0.03 (src~1)\\
17:49:31.77  & -28:08:05.4 & 24 & 22.20$\pm$0.08 & 21.94$\pm$0.12 &0.64 & 19.20$\pm$0.06 (src~2)\\
\end{tabular}
\end{center}
\end{table*}

Our star 1 at RA: 17:49:31.73 Dec: -28:08:05.09 would correspond to
source number 8 in table 2 of \citet{2011A&A...534A..92D} (see
Table~\ref{tab-avanzo}). Our star 2 which is at RA: 17:49:31.768 Dec:
-28:08:05.41 would correspond to their entry number 24. The
other four sources inside the \chan\, error circle are not
significantly detected in our {\it Gemini} data, although there is some
evidence for additional flux to the south of star 1, which could be
caused by either source 25 or source 27, or by both sources together.

The extinction towards \src\, in the $H$--band is $\sim$1.2 mag more
than that in the $K$--band. The $(H-K)_0$ colour of A--M5 stars is
between 0 and 0.3 mag (\citealt{2000asqu.book..143T}).  Therefore, the
magnitudes of the two sources that we find should be, at most, 1.5
magnitude brighter than the $H$--band magnitudes reported in
\citet{2011A&A...534A..92D} for these two sources.  However, comparing
our observed $K$ band magnitudes with the $H$--band magnitudes
reported by \citet{2011A&A...534A..92D} (see Table~\ref{tab-avanzo})
we see that the observed $H-K$ colour is approximately 2.4 magnitudes
for star 1 and 2.8 magnitudes for star 2. We investigated the
potential reason for this peculiar colour by comparing our magnitudes
as well as the $H$--band magnitudes derived by
\citet{2011A&A...534A..92D} with the magnitudes of 2MASS and UKIDSS
stars in the images. The $H$--band magnitude of the star at RA:
17:49:31.88 Dec: -28:08:03.3 reported by \citet{2011A&A...534A..92D}
(entry number 1 in their paper) is 16.26, however it has
$H$=15.01$\pm$0.05 according to the UKIDDS database, this is nearly
1.3 magnitudes brighter. In addition, we looked up the $H$ and $K$
magnitudes reported in the VISTA (Visible and Infrared Survey
Telescope for Astronomy) variable sources in the Via Lactea (VVV)
survey (\citealt{2010NewA...15..433M}, \citealt{2011rrls.conf..145C})
and the bright star has $H$=14.9, so close to the value provided by
UKIDSS. We conclude that the photometric calibration presented in
\citet{2011A&A...534A..92D} is off by 1.3 magnitudes. Indeed,
correcting for this photometric offset would imply observed $H-K$
colours of our stars 1 and 2 in line with what is expected given the
extinction towards the source. The alternative scenarios that could
potentially explain the observed $H-K$ colours, such as a
significantly higher extinction or a more shallow extinction law (a
large value of $R_V$) are not likely nor necessary to explain the
$H-K$ colours.

\begin{figure}\hbox{\hspace{-1.0cm}
\includegraphics[width=7cm,angle=90]{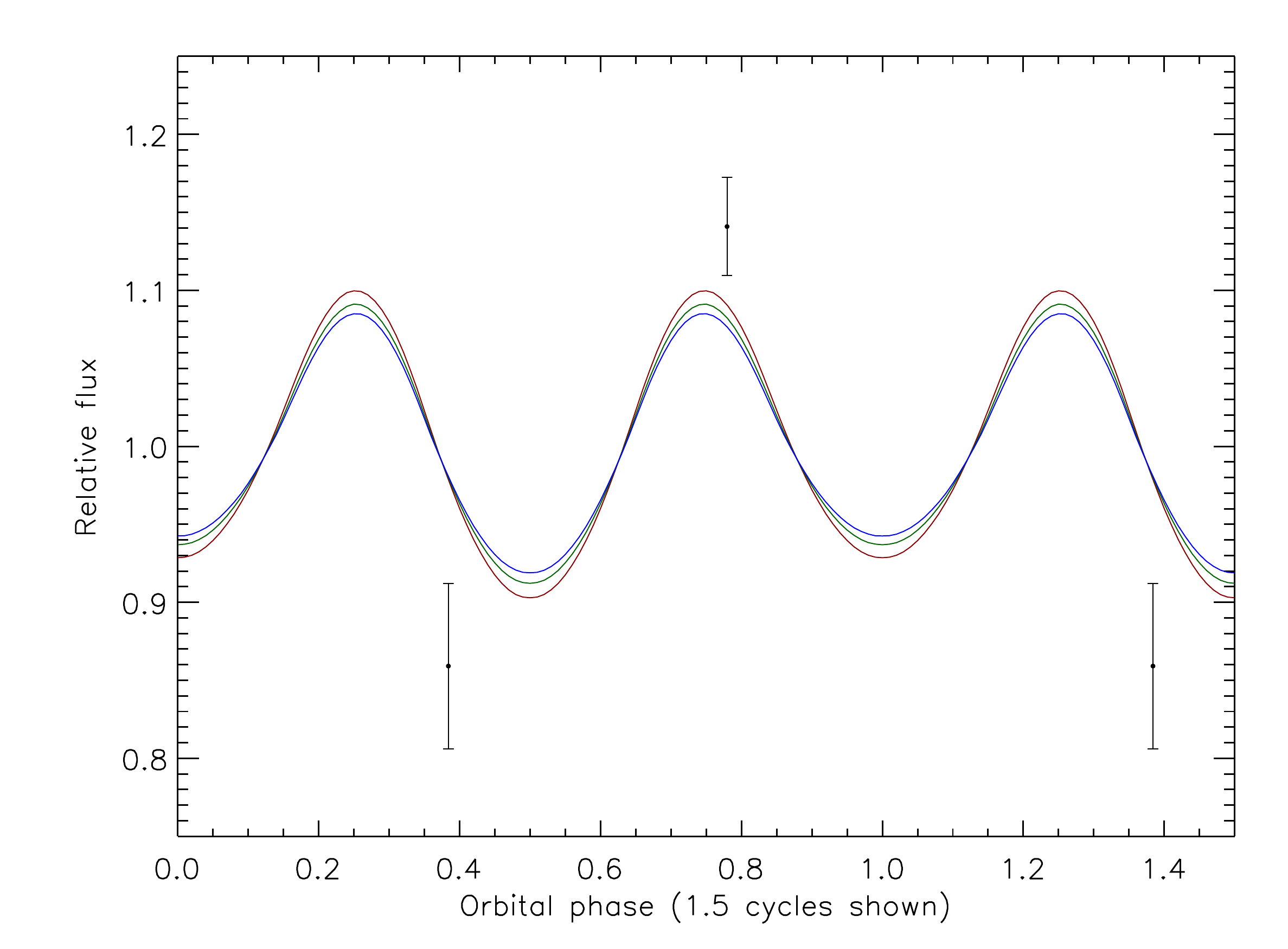}}
\caption{The expected ellipsoidal variations for a Roche lobe filling
  G5V star with in decreasing order of amplitude a mass ratio of
  $q=0.3,~ 0.5,~ 0.7$. The magnitudes for source 1 on the night of
  June 15, 2012 (near phase 0.77) and June 20, 2012 (near phase 0.36)
  are overplotted, showing that the observed variability has the right
  sign in the sense that the source is brighter near phase 0.77 than
  near phase 0.36. Changes in the predicted ellipsoidal modulations
  for a K0~V star are marginal with respect to those of the G5~V star
  shown.}\label{ellip}
\end{figure}

As a next step, spectroscopic follow-up is suggested to confirm/reject
source 1 as the counterpart. Spectra should show typical late type
star absorption features that should move as a function of the orbital
phase of \src. Typical features in the $K$--band spectra of late type
stars are the CO band head absorption bands
(e.g.~\citealt{2001A&A...373L..37G}; \citealt{2001Natur.414..522G}).
If confirmed, source 1 is within reach of current instrumentation for
time-resolved dynamical studies. If source 1 is found not to be the
counterpart, potentially, one can search for the right counterpart by
using adaptive optics observations under photometric conditions.
However, given that the maximum amplitude (peak--to--peak) of
ellipsoidal variations is 0.2 magnitudes for a mass ratio of $q=$0.3
this is challenging. In Figure~\ref{ellip} we plot the magnitudes of
source 1 over the expected ellipsoidal variations for a Roche lobe
filling companion star of spectral type G5V for three different values
of $q$. With decreasing amplitude of the modulations we plot $q=0.3,~
0.5,~ 0.7$. The figure shows that the observed variability has the
right sign given that the source is brighter near phase 0.77 than near
phase 0.36. For smaller mass ratios the peak--to--peak amplitude
increases, which given that the two data points lie above and below
the curves, could hint at a more extreme mass ratio. Such a mass ratio
could be accommodated by a neutron star more massive than 1.4~\msun~
and a mass donor star under--massive for its spectral type.
Furthermore, if light from a residual accretion disc is present it
will reduce the apparent amplitude of the ellipsoidal modulations.
Thus, adding a significant amount of light of an accretion disc to the
model can only be made in agreement with the two data points for a
(very) small mass ratio $q$.  Potentially, if the neutron star heats
the companion star one can find variations due to changes in the
aspect of the companion star (see for instance
\citealt{2001MNRAS.325.1471H}; \citealt{2003A&A...404L..43B};
\citealt{2008ApJ...680..615J}; \citealt{2009A&A...508..297D}).
However, such variations would possibly be larger than the ellipsoidal
variations and they would produce a bright $K$-band magnitude at
orbital phase of 0.5 and this is not seen in the current data (see
Figure~\ref{ellip}). Given that the orbital period of \src\, is larger
than that of the accretion powered millisecond pulsars where this
effect has been found before, the amplitude of the effect will be
reduced. Finally, flickering often observed in optical and
near--infrared light of X-ray binaries in quiescence due to for
instance short-term variations of the accretion disc can also explain
the observed variability (e.g.~ see \citealt{2008MNRAS.387..788R};
\citealt{2010ApJ...710.1127C}).

\section*{Acknowledgments} 

\noindent The authors acknowledge the referee for providing useful
comments that helped improve the manuscript. PGJ acknowledges support
from a VIDI grant from the Netherlands Organisation for Scientific
Research.  PGJ acknowledges Theo van Grunsven for discussions on the
amplitude of the ellipsoidal variations and for producing
Figure~\ref{ellip}. MAPT and PGJ acknowledge Knut Olsen for
discussions on the NIRI data reduction.  This research has made use of
data obtained from the Chandra Source Catalog, provided by the Chandra
X-ray Center (CXC) as part of the Chandra Data Archive
(ADS/Sa.CXO\#CSC).

\end{document}